\documentclass[sigconf]{acmart}

\AtBeginDocument{%
  \providecommand\BibTeX{{%
    \normalfont B\kern-0.5em{\scshape i\kern-0.25em b}\kern-0.8em\TeX}}}

\copyrightyear{2023} 
\acmYear{2023} 
\setcopyright{acmlicensed}\acmConference[WebSci '23]{15th ACM Web Science Conference 2023}{April 30-May 1, 2023}{Evanston, TX, USA}
\acmBooktitle{15th ACM Web Science Conference 2023 (WebSci '23), April 30-May 1, 2023, Evanston, TX, USA}
\acmPrice{15.00}
\acmDOI{10.1145/3578503.3583611}
\acmISBN{979-8-4007-0089-7/23/04}

\usepackage{multirow}
\usepackage{enumitem}
\usepackage{soul}

\begin{document}

\title[Emotional Framing in the Spreading of False and True Claims]{Emotional Framing in the Spreading of False and True Claims}


\author{Akram Sadat Hosseini}
\affiliation{%
  \institution{Institute for Parallel and Distributed Systems}
  \city{Stuttgart}
  \country{Germany}}
\email{akram.hosseini@ipvs.uni-stuttgart.de}

\author{Steffen Staab}
\orcid{1234-5678-9012}
\affiliation{%
  \institution{Institute for Parallel and Distributed Systems}
  \city{Stuttgart}
  \country{Germany}
}
\affiliation{%
  \institution{University of Southampton}
  \city{Southampton}
  \country{UK}
}
  \email{steffen.staab@ipvs.uni-stuttgart.de}

\renewcommand{\shortauthors}{Hosseini and Staab}

\begin{abstract}
  The explosive growth of online misinformation, such as false claims, has affected the social behavior of online users. In order to be persuasive and mislead the audience, false claims are made to trigger emotions in their audience. This paper contributes to understanding how misinformation in social media is shaped by investigating the emotional framing that authors of the claims try to create for their audience. We investigate how, firstly, the existence of emotional framing in the claims depends on the topic and credibility of the claims. Secondly, we explore how emotionally framed content triggers emotional response posts by social media users, and how emotions expressed in claims and corresponding users' response posts affect their sharing behavior on social media. Analysis of four data sets covering different topics (politics, health, Syrian war, and COVID-19) reveals that authors shape their claims depending on the topic area to pass targeted emotions to their audience. By analysing responses to claims, we show that the credibility of the claim influences the distribution of emotions that the claim incites in its audience. Moreover, our analysis shows that emotions expressed in the claims are repeated in the users' responses. Finally, the analysis of users' sharing behavior shows that negative emotional framing such as anger, fear, and sadness of false claims leads to more interaction among users than positive emotions. This analysis also reveals that in the claims that trigger happy responses, true claims result in more sharing compared to false claims.
\end{abstract}





\maketitle

\section{Introduction}\label{sec1}
The sheer volume of false claims published on the Internet has affected the social behavior of online users by manipulating public sentiment \cite{khaldarova2016fake}, influencing people’s perceptions, and distorting their awareness and decision-making \cite{zhang2019detecting}. False claims\footnote{In the current political climate the “fake news” term has been co-opted by politicians as a strategy for labeling news sources that do not support their positions \cite{golbeck2018fake}. Therefore, in this paper, we adopt the term “false claims” instead of using the term fake news.} comprise intentionally false statements of facts or knowingly fabricated statements to deceive users into believing they are true \cite{klein2017fake}. But what makes false claims so powerful? 

Let us consider the following tweet by Senator Ted Cruz, criticizing COVID-19 guidance urging vaccinated people to continue wearing masks, saying: \textit{“This is a bizarre, lunatic, totalitarian cult. It’s not about vaccines—it is instead profoundly anti-science, and is only focused on absolute govt control of every aspect of our lives."} This statement, widely shared in December 2020, has been rated as false in PolitiFact (\url{www.politifact.com}), a reputed US political fact-checking website. This claim tries to inspire strong emotions such as anger, fear, and sadness in the audience through emotionally loaded words like ‘bizarre’, ‘lunatic’, and ‘totalitarian’. 

\begin{figure}[t]
\centering
\includegraphics[width=\linewidth]{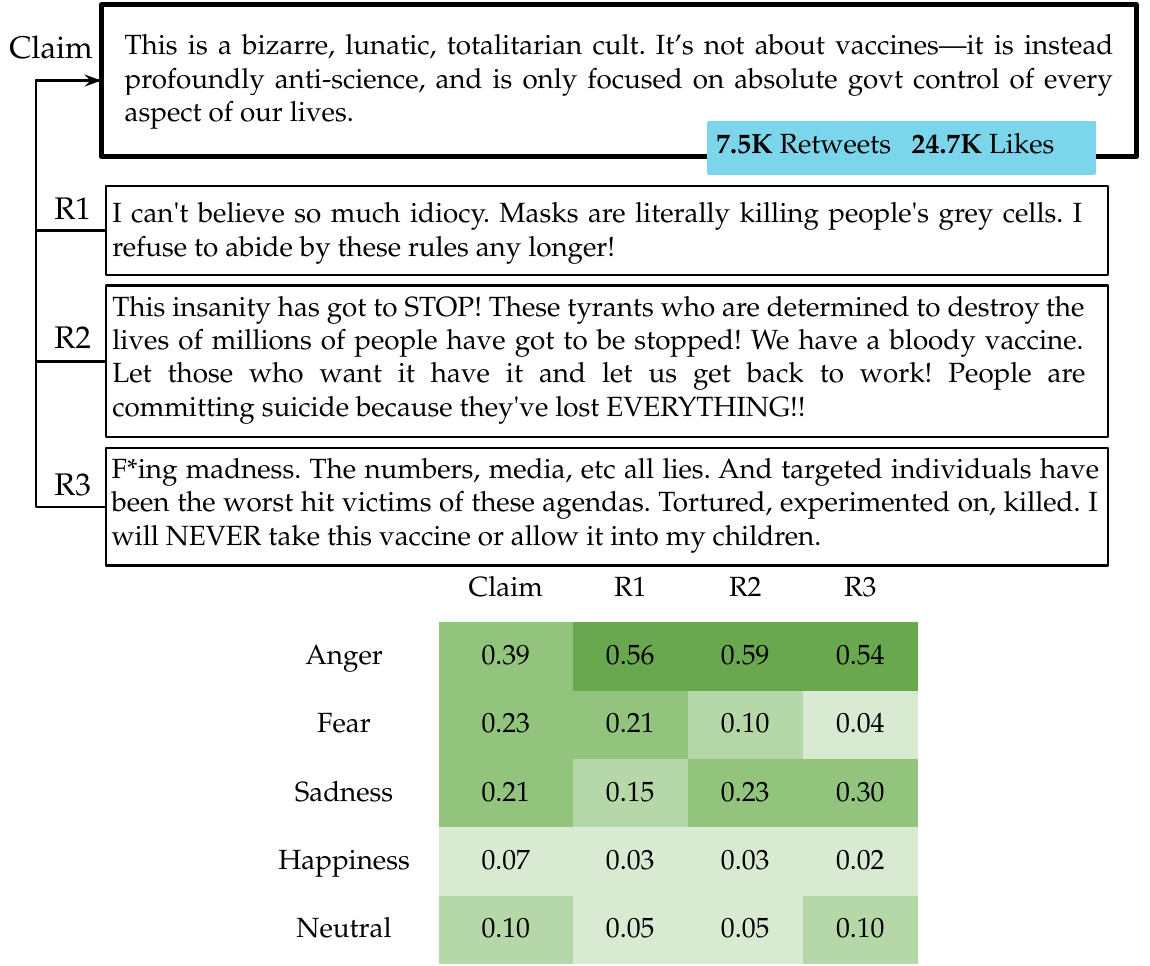}
\caption{An example of COVID-19 false claim and corresponding replies in Twitter with their emotional degrees.}\label{Example}
\end{figure}

Communication literature has shown that reading an emotionally loaded message makes readers rely on the emotions in the message to form their attitude \cite{kuhne2014political}. It seems that some power of the above statement is drawn from an {\ttfamily emotional framing} that sways its audience to accept the tweet as being true, to like and share the tweet, and to respond to it with likewise emotions (cf. Fig. \ref{Example}). Emotional framing is a way of presenting the content of a claim by using emotional words and sentences that influence the thinking or behavior of a claim's consumers, in particular, to direct and control discourse and attach a certain emotion to the audiences’ ideas and opinions about a given claim \cite{fish2019emotional}. Therefore, exploring emotional framing is key to understanding the impact of misinformation on the audience.

Research on false claims has been performed within two broad areas, namely, technical and behavioral \cite{suntwal2020does}. Technical research focuses on detecting the falsity of the information by using Natural Language Processing models and deceptive cues in the language of claims \cite{zhang2019detecting,wang2019learning}, while behavioral research focuses on human aspects like belief, attitude, intentions, etc to understand the impact of false claims on the formation of opinions \cite{suntwal2020does}. Our focus is on understanding how misinformation is shaped by exploring the impact of emotionally framed claims on their audience's behavior in social media. Therefore, we relate this work mostly to behavioral research.

Studies investigating the behavior of social media users with regard to misinformation showed that false claims are attention-grabbing \cite{osatuyi2018tale} and designed to make it hard for humans to identify them as false by exploiting people's cognitive limitations, emotions, and ideological biases \cite{sharma2019combating}. Using Elaboration Likelihood Model (ELM) \cite{osatuyi2018tale} theorizes two routes through which true and false content generators persuade readers: a central route of high cognitive effort, and a peripheral route of low cognitive effort. Their results indicate that false claims favor the peripheral route by providing less information and crafting information in a way that requires less cognitive effort from the recipient. In a recent study, \cite{vosoughi2018spread} revealed that falsehood diffused significantly further than the truth. They found that false claims inspire replies expressing intense emotions such as surprise and disgust. In a similar study, \cite{martel2020reliance} found a strong correlation between experiencing heightened emotions and an increased probability of belief in false news. Furthermore, \cite{su2020motivations, pellert2020individual, guo2019exploiting} posit that a negative tone was attractive to readers and therefore the authors of false claims emphasized strong negative emotions for persuading and misleading the audience to believe in falsified information and spurring them to act.  

While these studies highlight \textbf{that} emotions in false claims are important factors that impact people’s willingness to pass on information on social media, they fail to explain \textbf{how} emotions make false claims powerful such that they spread more widely than true claims. To answer this overall question, (i) we try to reproduce previous findings about negative emotions dominating false claims and try to generalize such findings over various topics. Then we turn to the effects that emotional false claims have on their audience, (ii), we study the audience’s attitude that false claims evoke by investigating response posts for uttered emotions. Finally, (iii), we link these attitudes to the audience’s behavior in terms of retweeting and liking claims, such that we better understand how emotions in false claims contribute to their broad spreading. 

Corresponding to (i), (ii), and (iii), we formalize the following research questions:

\begin{itemize}[leftmargin=*]
\item RQ1: In what way does the topic and credibility of a claim affect the emotional language used in a claim?
\item RQ2: How do emotions expressed in users’ response posts relate to emotions in false and true claims?
\item RQ3: How and to what extent do the emotions expressed in reply to falsehoods inspire activities such as retweeting and liking a claim?
\end{itemize}

To operationalize the discovery of emotions in utterances to the fullest possible extent, we develop a computational framework, based on DepecheMood ++ \cite{araque2019depechemood++}, NRC Affect \cite{mohammad2018understanding}, and NRC Valence \cite{mohammad2018obtaining} lexicons, for measuring emotions in four emotional categories (anger, fear, sadness, and happiness). Our method unifies these lexical resources which are based on different emotion representations in a joint representation of four emotion labels, i.e. anger, fear, sadness, and happiness (more details in Section \ref{sec4}). 

Our analysis of four real-world data sets covering different topics (politics, health, Syrian war, and COVID-19) showcases the interdependency between the topic of the claim and the emotions expressed in the claims, as well as, the influence of the claims’ credibility on the strength of emotions expressed in the claims, rebuking previous findings that false claims would always focus on negative emotions. Analyzing users’ emotional responses on social media to claims reveals that emotions expressed in the claims are repeated in the users' responses. Moreover, the results indicate the reciprocal effect of the claims' credibility on users' sharing behavior. Our analysis reveals negative emotions such as anger, fear, and sadness at false claims lead to more interaction among users than positive emotions. However, interestingly, in the claims expressing happy emotions, true claims triggered a higher level of interaction between users than false claims.

The rest of the paper is organized as follows: First, an overview of the related work is presented. Next, the proposed research model and hypotheses are described. Then, the research methodology including the implementation details is presented. Afterward, we discuss data analysis and results, and finally, we draw conclusions.

\section{Literature Review}\label{sec2}
This section describes relevant research on misinformation that has been grouped into three broad areas, namely, content-, social-, and behavioral-level analysis.

\subsection{Content Level Analysis}
Early research on misinformation has been focused on analyzing writing styles and content of news from a linguistic perspective to develop better detection methods \cite{suntwal2020does}. False claims mimic legitimate claims by imitating language styles of true claims or expressing opinions with a tone frequently used in true claims \cite{shu2020hierarchical}. However, “language leakage” or so-called “deception cues” occur with certain verbal aspects in the content of a claim that is hard to monitor \cite{golbeck2018fake}. For example, a fine-grained analysis of word usage in deceptive texts by \cite{larcker2012detecting} revealed that deceivers are likely to use more other-oriented pronouns (you, your), instead of self-oriented pronouns (I, me), indicative of the speaker’s discomfort in identifying themselves with the lying statements. In a similar study, \cite{mihalcea2009lie, newman2003lying} indicated that negative emotion verbs (e.g., hate, worthless, envy) and words related to “certainty” are dominant in deceptive texts, whereas, true claims contain significantly fewer hedges, subjective terms, and harmful words than false claims \cite{volkova2017separating}. In a recent study, \cite{mackeydetecting} demonstrated that combining different emotional features extracted from news content with word-embedding models, such as BERT, can enhance the accuracy of classifying misinformation.

\subsection{Social Level Analysis}
Social contextual features have been used in various linguistic-based approaches to enhance veracity detection \cite{shu2019beyond}. Using a stance-based model, \cite{jin2016news} defined a scheme to categorize types of reactions expressed toward false and true claims to infer news veracity. In a similar study, \cite{glenski2018identifying} showed that posts linked to deceptive sources receive more questions, appreciation, and answers from users, while posts linked to trusted sources have higher rates of elaboration, agreement, and disagreement. Furthermore, social psychology research indicates that emotion-eliciting content influences the sentiment of readers \cite{van2017nature, wang2019emotional}. Consequently, \cite{wu2020adaptive, zhang2021mining} emphasized the importance of considering not only the emotions expressed in the content of news but also the emotions elicited in social media users when faced with the news, for the purpose of detecting its veracity. Moreover, according to \cite{zhang2021mining}, there are more instances of false news articles in which both the claim and the user response posts express anger, compared to real news. Arguing that users' interaction with information depends on their interest in its topic, \cite{roberts2012empatweet} showed users express different emotions towards 10 different Twitter topics. In a similar study, \cite{del2016spreading} reports that while misinformation around scientific news reaches a higher level of diffusion faster, it also decays faster.

\subsection{Behavioral Level Analysis}
Increasingly, false news generators successfully persuade recipients to change their attitudinal responses in the desired direction. This has led recent research to behavioral analysis \cite{jia2017rumors}. The behavioral analysis investigates social aspects like belief, attitude, intention, etc to explain individual behaviors with regard to misinformation \cite{suntwal2020does}. Using reputation theory, \cite{kim2018says} found that confirmation bias, source ratings, and argument quality, which are moderated by the veracity of information, had a significant positive effect on believability, and believability had a further positive effect on activities like commenting, and sharing. By exploiting the emotional content of tweets, \cite{katsyri2016negativity} found that individuals are more likely to engage with negative information. This is supported by research that suggests that information reflecting negative emotions such as anger, fear, sadness, or doubt are spread more readily from person to person and more widely through social media than posts expressing joy \cite{pellert2020individual}. Further research has shown that negativity boosts the likelihood of one’s content being shared, particularly more for verified users on Twitter than ordinary users \cite{schone2022negative}. Additionally, an examination of the relationship between experiencing emotions and believing false news found that heightened emotionality is associated with an increased likelihood of belief in fake news and a decreased ability to differentiate between True and false news posts \cite{martel2020reliance}.

Although the existing studies have revealed the significance of emotions in false claims and their influence on individuals' inclination to pass on information on social media, they fail to explain how these emotions fuel the greater spread of false claims compared to true ones. This paper contributes stronger empirical evidence compared to prior research in order to gain a comprehensive understanding of the role of emotions in the spread of false claims. Building on prior findings that suggest negative emotions are often associated with false claims, our research seeks to generalize these conclusions across various topics. Moreover, we analyze audience response posts to these claims to gauge their emotional reactions, and finally, we assess the influence of such emotions on their sharing behavior.


\section{Research Model}\label{sec3}

Building upon previous studies presented in the literature review, to yield an interpretable analysis of information sharing behavior in social media, we set out a three-level research model underlying our study in Fig. \ref{research_model}. Details about the phases sketched in the research model have been explained in the following:

\begin{figure}[t]
      \centering
  \includegraphics[width=\linewidth]{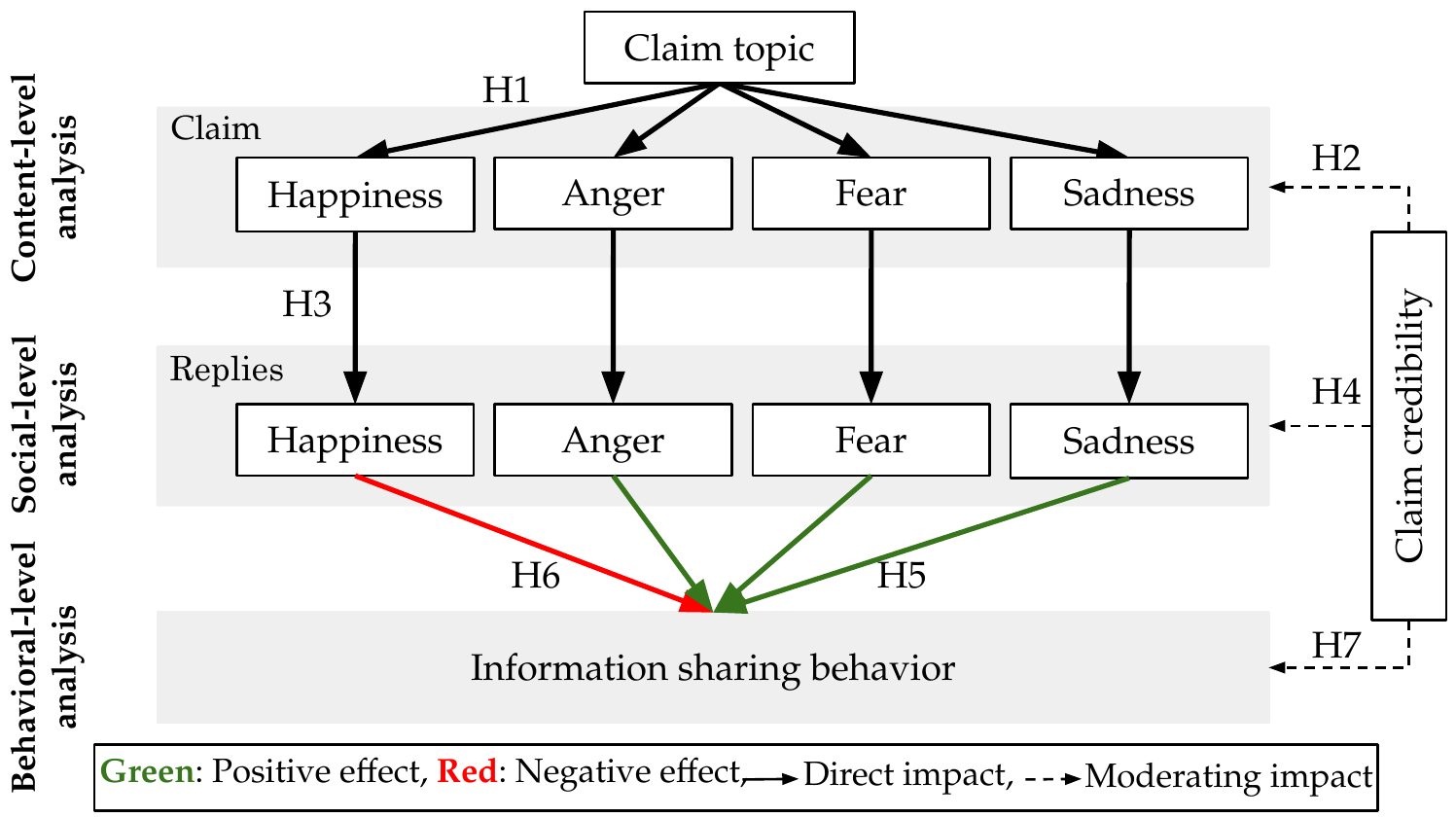}
  \caption{Proposed Research Model}\label{research_model}
\end{figure}

\noindent\textbf{Content-level Analysis.} At the first level of analysis, our goal is to understand how misinformation is expressed from an emotional perspective. For this purpose, we explore the correlation of emotions expressed in the claims with the claims’ topic and credibility. To this end, the hypotheses developed are as follows:

\begin{itemize}[leftmargin=*]
    \item\textbf{H1.} The topic of a claim has a direct impact on the emotional language used in a claim.
    \item\textbf{H2.} Credibility of the claim has a moderating effect on the strength of emotions expressed in a claim.
\end{itemize}

\noindent\textbf{Social-level Analysis.} At the second level of analysis, we investigate how the emotional framing of topics in claims with different credibility influences the emotion of response posts by social media users to the claims. Indeed, we want to observe the reactions of readers by studying the emotions that they express in their responses. The underlying foundation of our idea is based on the assumption that the main emotion in the claims has a positive direct effect on the emotions of users expressed in their posts responding to these claims. The above arguments lead to the following hypotheses:

\begin{itemize}[leftmargin=*]
    \item \textbf{H3.} The level of dominant emotion (happiness, anger, fear, and sadness) in a claim has a direct effect on the level of users’ emotions expressed in their posts responding to the claim.
    \item\textbf{H4.} A claim's credibility has a moderating effect on emotions of response posts.
\end{itemize}

\noindent\textbf{Behavioral-level Analysis.} Finally, the third level of analysis investigates the influence of users' emotional responses related to false and true claims on sharing behavior in social media. The underlying assumption at this level of analysis is that, along with the topic, the emotion that was aroused in users towards a topic is an important factor in modeling activities such as retweeting and liking a claim. Thus, we explore the following hypotheses:

\begin{itemize}[leftmargin=*]
    \item\textbf{H5-6} Claims that trigger high levels of anger, fear, and sadness among users have a higher potential to be shared on social media as compared to claims that trigger a high level of happiness.
    \item\textbf{H7.} The credibility of the claims has a moderating effect on the sharing level of the questionable claims in social media.
\end{itemize}

\section{Measuring Emotions Methodology}\label{sec4}

\begin{figure}[t]
      \centering
  \includegraphics[width=\linewidth]{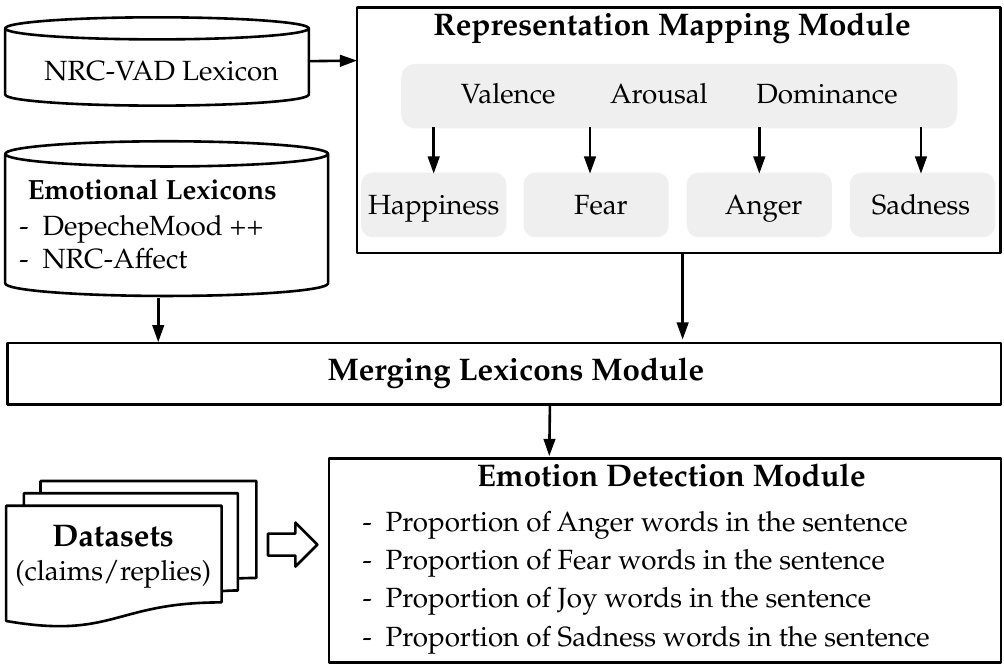}
  \caption{Overview of the framework for measuring emotions in claims and responses}\label{methodology}
\end{figure}

The computational framework of our methodology for measuring emotions is illustrated in Fig. \ref{methodology}. It includes three main modules, named Representation Mapping Module, Merging Lexicons Module, and Emotion Detection Module. The Representation Mapping and Merging Lexicons modules describe the approach that creates a unified emotion lexicon with high word coverage, enriched with the emotional intensity of words. The NRC Affect lexicon \cite{mohammad2018understanding} serves as the basis for the joint representation. This lexicon was manually annotated with four basic emotions - \textit{anger, fear, sadness, joy} - for both common English terms and terms that are prominent on social media platforms. The DepecheMood++ \cite{araque2019depechemood++} and NRC-VAD \cite{mohammad2018obtaining} lexica are mapped into the representation used by NRC Affect, and the results are merged. The Emotion Detection Module extracts emotion vectors using the merged lexicon. The next sections provide details about the phases sketched in Fig. \ref{methodology}.

\subsection{Representation Mapping Module}
The presence of emotional words in a sentence provides a good premise for the interpretation of the overall emotion of the sentence \cite{hosseini2017sentence}. Therefore, our first step towards emotion detection is discovering keywords and phrases that associate with emotion. Table \ref{lexicons} summarizes well-accepted resources used for acquiring basic knowledge about emotion-carrier words.

\begin{table*}[t]
\begin{center}
\caption{Summary of the computational resources used in this paper.}\label{lexicons}
\tabcolsep=0.09cm
\resizebox{\textwidth}{!}{%
\begin{tabular}{lp{15.5cm}}
\toprule
{Lexicon} & {Description} \\  
\midrule
 DepecheMood ++ &  DepecheMood ++ \cite{araque2019depechemood++} is a high-precision and high-coverage emotion lexicon that has been automatically derived from crowd annotated news. It provides eight values between 0 and 1 (\begin{math}e \in [0, 1]^8\end{math} and $||e||=1$) for 62,224 entries on the following dimensions: \textit{anger, anticipation, disgust, fear, joy, sadness, surprise,} and \textit{trust}. \\  
 NRC-Affect &  NRC Affect Intensity lexicon \cite{mohammad2018understanding} is a list of 4,192 English words and their associations with four basic emotions (\textit{anger, fear, sadness, joy}) with scores ranging from 0 to 1 (\begin{math}e \in [0, 1]^4\end{math}). It includes common English terms as well as terms that are prominent in social media platforms. \\
 NRC-VAD  &  NRC Valence (positive--negative), Arousal (excited--calm), and Dominance (powerful--weak) lexicon \cite{mohammad2018obtaining} includes a list of more than 20,000 English words mapped to a 3D vector of VAD values, ranging from 0 (lowest) to 1 (highest) (\begin{math}e \in [0, 1]^3\end{math}).\\
\bottomrule
\end{tabular}}
\end{center}
\end{table*}

We investigated how similar to each other and different from each other these resources are. Table \ref{uniqueness&intersection} shows the number of common words between each pair of lexicons, and also the number of words from each lexicon that have not been found in the remaining lexicons. The high number of unique words indicates that high coverage of emotional words can best be achieved by integrating lexicons. However, their integration is non-trivial, as the different resources are based on different representations. For example, the representation of the word \textit{deadly} appears in different ways in the three lexicons (see Table \ref{representations}). In DepecheMood++ and NRC-Affect \textit{fear} has the highest score, and in the NRC-VAD lexicon, arousal has the highest score.  

\begin{table*}[t]
\begin{center}
\caption{Number of unique words in each lexicon and number of words that overlap between each pair of lexicons.}\label{uniqueness&intersection}
\begin{tabular}{lccccc} 
\toprule
\multirow{2}{*}{{Lexical resource}} & \multirow{2}{*}{{\#Words}} & \multirow{2}{*}{{\#Unique words}} & \multicolumn{3}{@{}c@{}}{{\#Common words}}  \\ 
\cmidrule{4-6}
&  &   & DepecheMood ++ & NRC-Affect & NRC-VAD  \\ 
\midrule
DepecheMood ++ & 62,224 & 54,456 & - & 1,790 & 7,759\\ 
\cline{1-6}
NRC-Affect  & 4,192 & 0 & 1,790 & -  & 4,192 \\ 
\cline{1-6}
NRC-VAD & 20,006 & 9,740 & 7,759 & 4,192 & - \\
\bottomrule
\end{tabular}
\end{center}
\end{table*}

\begin{table}[t]
\begin{center}
\caption{Representation of the word \textit{deadly} in lexicons.}\label{representations} 
\begin{tabular}{ll}
\toprule
Lexical resource & Representations \\ 
\midrule
DepecheMood++ &  Anger \hspace{10pt}         Fear \hspace{10pt}          Sadness  \hspace{10pt}         Happiness \\
&  0.17 \hspace{18pt}   0.36 \hspace{11pt}         0.21 \hspace{24pt} 0 \\  
\cline{1-2}
NRC-Affect &  Anger \hspace{10pt}         Fear \hspace{10pt}          Sadness  \hspace{10pt}         Happiness \\
& 0.76 \hspace{18pt}       0.90 \hspace{11pt}       0.88 \hspace{24pt}     0 \\
\cline{1-2}
NRC-VAD  &  Valence  \hspace{21pt}        Arousal    \hspace{20pt}       Dominance \\
& 0.14 \hspace{35pt}  0.85  \hspace{34pt}   0.55 \\
\bottomrule
\end{tabular}
\end{center}
\end{table} 

Exploring different lexicons revealed the need for a unified representation. To deal with different representations, we have developed the representation mapping module that maps the VAD representations of words in the NRC-VAD lexicon into ratings for four emotion labels (i.e. anger, fear, sadness, happiness), which is a joint representation covered by both the NRC-Affect and DepecheMood++ lexicons. According to Russell and Mehrabian \cite{russell1977evidence} discrete affective states can be represented by the VAD dimensions. Table \ref{Mapping} presents the corresponding values of the VAD model for \textit{happiness, anger, fear} and \textit{sadness} emotions. Employing Table \ref{Mapping}, we determined a four-dimensional emotional model of NRC-VAD lexicon words by computing the cosine similarity between the words and \textit{happiness, anger, fear} and \textit{sadness} emotions, as shown in \eqref{Similarity},

\begin{align} \label{Similarity}
    \mbox{Emotional Score} (\mbox{w}, \mbox{i}) = \frac{ \mbox{VAD}_{\mbox{w}} \times \mbox{VAD}_{\mbox{i}}}{\|\mbox{VAD}_{\mbox{w}}\|\|\mbox{VAD}_{\mbox{i}}\|} \\\mbox{i} \in\big\{\mbox{Happiness, Anger, Fear, Sadness}\big\} \nonumber
\end{align} 

This equation assigns an emotional score to every word \textit{w} in the NRC-VAD lexicon based on four emotional dimensions of \textit{happiness, anger, sadness,} and \textit{fear}. To avoid different label spaces that exist in different lexicons, before merging lexicons, values have been rescaled linearly such that the simple sum of them equals 1.

\begin{table}[t]
\begin{center}
\caption{Values for \textit{happiness, anger, fear} and \textit{sadness} emotions in terms of VAD emotion dimensions \cite{russell1977evidence}.}\label{Mapping} 
\begin{tabular}{lccc}
\toprule
Emotion & Valence & Arousal & Dominance \\ 
\midrule
Happiness        & 0.76             & 0.48             & 0.35               \\ 
Anger            & -0.51            & 0.59             & 0.25               \\ 
Fear             & -0.64            & 0.60             & -0.43              \\ 
Sadness          & -0.63            & -0.27            & -0.33              \\ 
\bottomrule
\end{tabular}
\end{center}
\end{table}

\subsection{Merging Lexicons Module}\label{sec4.2}
To achieve maximum coverage of emotion carrying words, the Merging Lexicons Module integrates NRC-Affect, DepecheMood++ and NRC-VAD into our joint lexicon. NRC-Affect constitutes the basis to which words from DepecheMood++ are added if they are not available in NRC-Affect. To add a word from DepecheMood++ its 8-dimensional representation is shortened to consider only its happiness (=joy), fear, anger, and sadness scores. Finally, words that are in NRC-VAD, but not yet represented, are added based on the mapping from equation \ref{Similarity}.

\subsection{Emotion Detection Module}
Emotion detection refers to the task of automatically assigning an emotion score to text from a set of predefined emotion labels \cite{alswaidan2020survey}. Emotion lexicons are commonly used resources in unsupervised techniques to automatically and straightforwardly label text with emotional information, known as the keyword-based approach \cite{baziotis2018ntua, ma2018targeted, cambria2017practical}. To discover emotions, we employ the unified emotion lexicon described in Section \ref{sec4.2}, which is based on DepecheMood++, NRC Affect, and NRC-VAD lexical resources. The "Emotion Detection Module" relies on the statistical features of emotional words to produce a vector of real-valued numbers as the result of the emotion assessment.

In the pre-processing phase, non-semantic words such as prepositions, conjunctions, and pronouns also known as stop words are removed from the input text. For out-of-vocabulary words, we used a difflib library in python to find the closest word to a given OOV word in the final joint lexicon. If no word with more than 90\% of similarity is found, the word is determined as neutral with a default value \cite{witon2018disney}. We define proportions of happiness, anger, fear, sadness, or neutrality by extending our four-dimensional model to a five-dimensional. Now, given a sequence of words $s=(w_1 ...w_k)$ we can calculate the proportion for any of the five dimensions by

\begin{align} \label{MetaEmotion}
\mbox{Proportion (s, i)} = \frac{\mbox{1}}{\mbox{k}} \sum_{j=1}^k  \pi_i (w_j)\\\mbox{i} \in\big\{\mbox{Happiness, Anger, Fear, Sadness, Neutral}\big\} \nonumber
\end{align} 

where $\pi_i$ is the projection of the j-th component for emotion $i$. To better interpret the measured values in formula \eqref{MetaEmotion}, we normalized emotional values according to the corresponding vector.  Indeed, each value in the vector is located between 0 and 1, and the sum of the values for a sentence is equal to 1. Fig. \ref{Example} shows an example of emotion measurement using statistical features of emotional words.


\section{Data Analysis and Results}

In this section, we provide an assessment of our hypotheses on four recently released real-world data sets. Data analysis was performed using Structural Equation Modelling (SEM) with SmartPLS v3.3.9. The reason we used Partial Least Squares (PLS) was related to the research model complexity with moderating variables \cite{hair2019factors}. In addition, PLS accommodates non-normally distributed data. Our measurements were not normally distributed, both skewness and kurtosis were significant, which indicates that PLS is suited to this study \cite{hair2017mirror}. Following the two-step analytical procedure suggested by \cite{hair2019factors}, to estimate and analyze path coefficients, and test the proposed research hypotheses, we first assessed the measurement model, then, performed structural model analysis.

\subsection{Data sets}
The challenging problem of analyzing false claims is confined so far to one particular domain mostly with a focus on the field of politics. It was our objective to investigate the influence of topics on the emotional framing of the claims therefore we selected data to cover different topics. Secondly, data sets must contain the annotated claims as false and true along with users' responses to the claims. We evaluate our hypotheses across 4 recently released real-world data sets from four topics: politics, health, Syrian war, and COVID-19. To have a reliable analysis of results, in all data sets we selected a balanced number of false and true claims. Table \ref{datasets} summarizes details of data sets used in this paper.

\begin{table}[t]
\begin{center}
\caption{Data sets characteristics}\label{datasets} 
\begin{tabular}{lccc}
\toprule
Datasets & Domain &  \#Claims  & \#Replies\\ 
\midrule
FA-KES \cite{salem2019fa} & Syrian War & 752  & - \\ FakeHealth \cite{dai2020ginger}& Health & 1,366 & -\\ 
\cline{1-4}
CoAID \cite{cui2020coaid} & COVID-19  & 1,892  & 47,547\\   
Co-inform \cite{denaux2021weaving} & Political  & 884   & 48,349\\ 
\bottomrule
\end{tabular}
\end{center}
\end{table}

\begin{itemize}[leftmargin=*]

    \item\textbf{FA-KES} \cite{salem2019fa} data set comprises claims labeled as true or fake that cover the many facets of the Syrian war from the year 2013 to 2017. The annotation was supported by employing a semi-supervised fact-checking approach with the help of crowd-sourcing. To validate the accuracy of the labeling approach, a set of 50 claims was also manually assessed by domain experts.

    \item\textbf{FakeHealth} \cite{dai2020ginger} data set is collected from 2005 to 2018 from Health News Review (\url{www.healthnewsreview.org}) website which assesses health news in aspects such as overclaiming, missing information, reliability of sources, and conflict of interests. The rating ranges from 0 to 5, and it is in proportion to the number of criteria satisfied by the claim. Following the strategy in \cite{shu2020fakenewsnet}, claims whose scores were lower than 3 were considered false.

    \item\textbf{CoAID} \cite{cui2020coaid} (\textbf{C}ovid-19 he\textbf{A}lthcare m\textbf{I}sinformation \textbf{D}ata set) includes COVID-19 related claims, ranging from December 2019 through November 2020. Annotation has been made through fact-checking websites where the credibility of the claims is assessed by experts, which makes the quality of annotation high. Users’ social engagement was also collected from Twitter which included the tweets discussing the claim in question and replies. 

    \item\textbf{Co-inform} \cite{denaux2021weaving} data set contains political claims collected during the year 2020 from different fact-checking websites - such as PolitiFact, Factcheckni (\url{www.factcheckni.org}), FullFact (\url{www.fullfact.org}) and labeled as credible, when the fact-checkers detect it as true, or not-credible when the fact-checkers detect the claim is false. The collected claims in this data set check if the claim appeared on Twitter. This filter helped us to collect social context related to the claims from Twitter.
\end{itemize}

FA-KES and FakeHealth data sets contain only annotated claims as false and true, but, CoAID and Co-inform data sets in addition to the claims contain also responses, like, and retweets related to the claims.

\subsection{Measurement Model Analysis}
We performed measurement model validation, using SmartPLS v3.3.9, on data sets to assess the reliability and validity (convergent and discriminant validity) of constructs before testing hypotheses. 

To this end, first, the \textit{measurement reliability} is verified using composite reliability (CR) and Cronbach’s alpha (CA) for each of the constructs \cite{chin1998partial, clark2016constructing}. Both values exceeded the recommended threshold value of 0.7 \cite{hulland1999use} for all constructs, which can thus be regarded as reliable. Second, the \textit{convergent validity} of constructs was assessed using the average variance extracted (AVE) \cite{fornell1981evaluating}. As expected, all constructs showed satisfactory values above the suggested threshold value of 0.50. Thus, our study met the requirement for convergent validity. Third, to determine the \textit{discriminant validity} of the measurement model, we assessed the Fornell-Larcker criterion of the constructs \cite{fornell1981evaluating}. Since the square root of AVE for each construct exceeded the corresponding inter-construct correlations, thereby, the discriminant validity of the constructs for the constructs was supported. Overall, the results demonstrated the high reliability and validity of the measurement model.

\subsection{Structural Model Analysis}
This section proceeds to test the proposed research model. The PLS-SEM approach with a bootstrapping procedure was used to assess hypotheses. The key criteria for evaluating the structural model and the strengths of the relationships between the variables are to examine the $t$-values, and path coefficient ($\beta$ values) \cite{hair2017primer}. The following sections provide the results of hypotheses testing on content-, social-, and behavioral-level.

\subsubsection{Content-level Analysis}
This section evaluates hypotheses H1 and H2 at the content-level of our research model, particularly, whether claims with different credibility elicit emotions with similar distributions on different topics. Our goal, in the first-level of analysis, is to understand if the basic word usage patterns differed from an emotional perspective in false and true claims and in different topics.

\begin{figure*}[t]
    \centering
    \includegraphics[width=\textwidth]{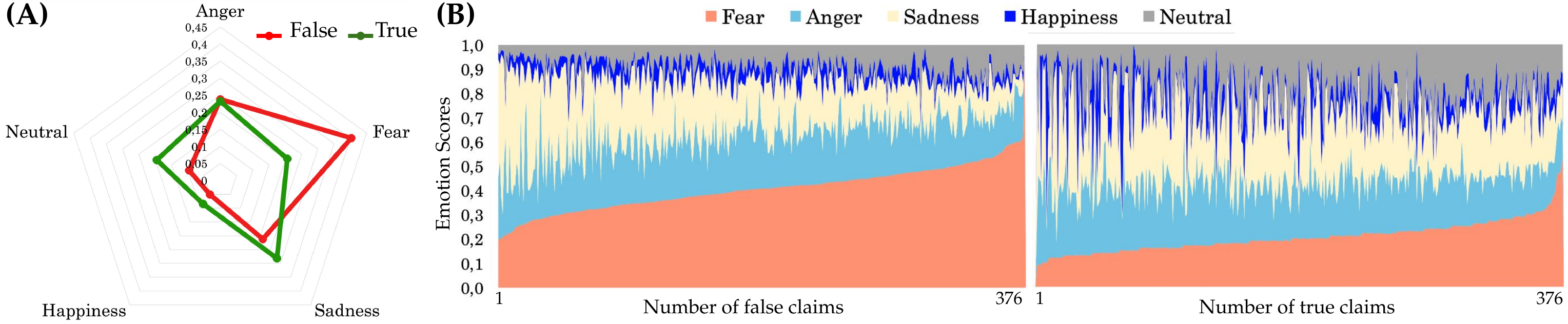}
    \caption{The averaged emotion scores (A) and emotional pattern diagrams (B) in FA-KES data set.}\label{FA-KES}
\end{figure*}

Fig. \ref{FA-KES} provides a graphical comparison of emotions expressed in the FA-KES data set. Chart A  shows the average of anger, fear, sadness, happiness, and neutrality in true and false claims. As one can see, in general, the claims convey negative emotions to their audience. Meanwhile, false claims show a higher intensity of “fear” compared to true claims, which makes sense since the topic of the claims is war. To have a better overview of emotions distribution in this data set, chart B depicts the emotional patterns in true and false claims. We call the distribution of scores over our four different emotions and neutrality the emotional pattern of a claim. The x-axis shows false and true claims and the y-axis shows the level of different emotions in claims. False and true claims demonstrate different emotional patterns. In true claims, you can see a similar distribution of emotion scores between anger, fear, and sadness. True claims tend to be more neutral than false claims, however, false claims tend to convey fear. Therefore, using the emotional pattern diagram it is predictable that, in the war-related claims, the level of fear expressed in claims is a signal that helps to assess their credibility.

\begin{figure*}[t]
    \centering
    \includegraphics[width=\textwidth]{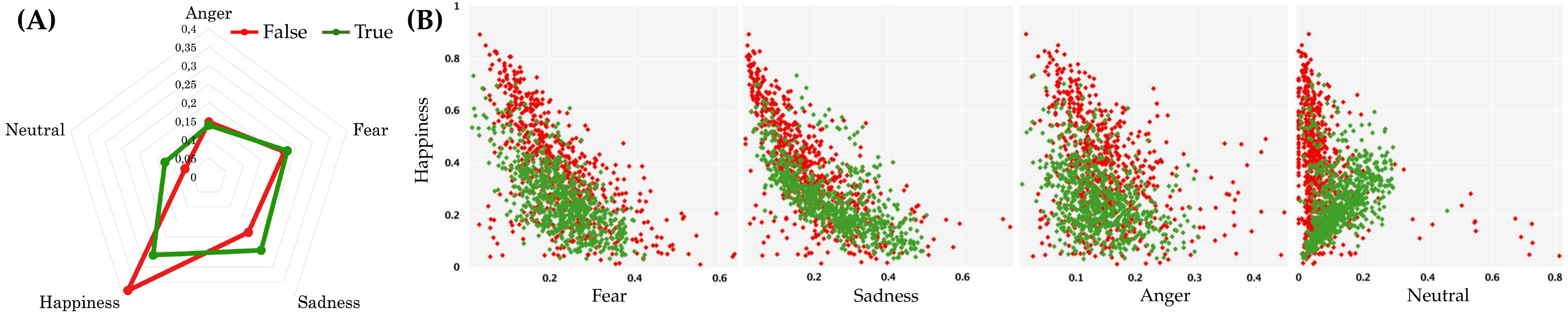}
    \caption{The averaged emotion scores (A) and Scatter plots of happiness intensity vs fear, sadness, anger, and neutral (B) in FakeHealth data set.}\label{FakeHealth}
\end{figure*}

Fig. \ref{FakeHealth} shows the correlation of emotions expressed in claims across the FakeHealth data set. Chart A  shows the average score of different emotions in the claims. We can see that there is a balance between negative emotions in false and true claims, while the average of “happiness” expressed in false claims compared to true claims is higher. Chart B shows the scatter diagram that compares happiness with fear, sadness, anger, and neutrality in claims. Here it is more clear that happiness has higher intensity compared to other emotions in false claims. Detailed observations have shown that most of the false claims transmit intense “happiness” by promising an outstanding medicine or approach to cure diseases like cancer and diabetes, which is understandable since companies try to advertise and make a sale of their products.

\begin{figure}[t]
  \centering
  \includegraphics[width=\linewidth]{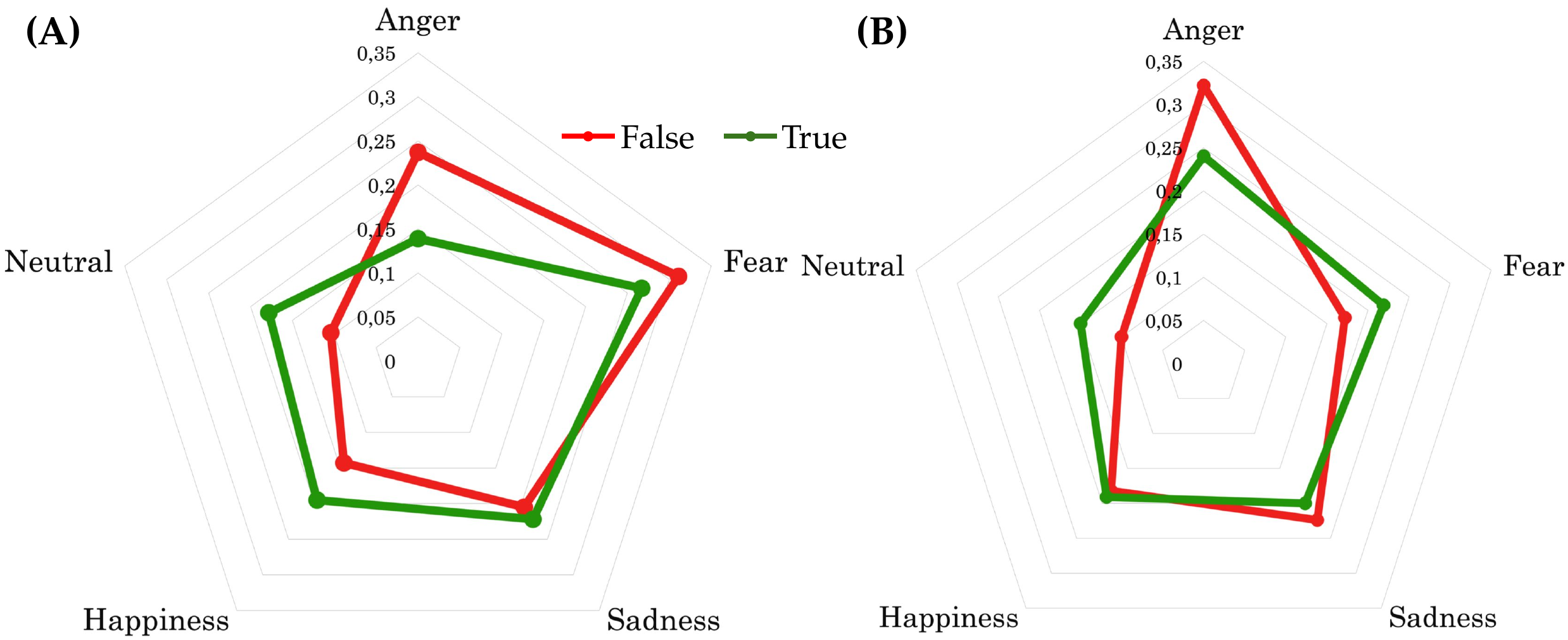}
  \caption{The average of emotion scores across true and false claims in CoAID (A) and Co-inform (B) data sets.}\label{CoAID-Coinform-Claims}
\end{figure}

Fig. \ref{CoAID-Coinform-Claims} shows the average of different emotions expressed in false and true claims in the CoAID (chart A) and Co-inform (chart B) data sets. 
In true claims related to COVID-19, there is a balance among emotions, as well as a tendency for neutrality compared to false claims. While, in false claims, an intense inclination towards fear and then anger can be observed. 
In the Co-inform data set, containing US political claims, as we can see, regardless of the claims' credibility, negative emotions (i.e. anger, sadness, and fear) are dominant. Although in false claims the intensity of anger is higher than true claims.

Table \ref{Hyp1} illustrates the path coefficient, $t$-values, and significance, for hypothesis H1. Hypothesis H1 addresses the structural relationships between the topic of a claim and the emotion expressed in a claim. To evaluate hypothesis H1, we combined claims from the four data sets and considered the topic of each data set as the topic of claims. As hypothesized, results reveal that the topic of the claim significantly influences the emotional language used in the claim. As is evident, the topic of the claims significantly affects the level of anger ($\beta = 0.37$; $p < 0.001$), fear ($\beta = 0.25$; $p < 0.001$), and sadness ($\beta = 0.17$; $p < 0.01$) expressed in the claims. Overall, this provides support for hypothesis H1. The relationship between the topic of the claim and the level of happiness expressed in the claims was not found significant ($p > 0.05$). 

\begin{table}[t]
\begin{center}
\caption{Structural model results for hypothesis 1.}\label{Hyp1} 
\begin{tabular}{@{}ll@{}}
\toprule
Hypothesised Relationship &  Path Coefficient($t$-value) \\ 
\midrule
Claim Topic $\,\to\,$ Claim Anger & 0.37 (5.16)***\\ 
Claim Topic $\,\to\,$ Claim Fear & 0.25 (4.53)*** \\ 
Claim Topic $\,\to\,$ Claim Sadness & 0.17 (2.95)**\\ 
Claim Topic $\,\to\,$ Claim Happiness & 0.01 (0.88)ns\\ 
Claim Topic $\,\to\,$ Claim Neutral &  0.05 (1.22)ns\\ 
\bottomrule
\end{tabular}
\small{Note: * Significant at $p<0.05$, ** at $p<0.01$, and *** at $p<0.001$.}
\end{center}
\end{table}

The path coefficient, $t$-values, and significance, for the hypothesis H2 testing results in each data set, are presented in Table \ref{Hyp2}. The results confirm that the credibility of the claims has a significant effect on the emotions expressed in the claims, offering support for H2. For example, in the FA-KES data set, containing claims related to the Syrian war, the credibility of the claims illustrates a positive direct effect on emotions, with the exception of fear ($\beta = -0.79$; $p < 0.001$). This means that the claims' credibility has a reverse significant effect on the level of fear in the claims, approving our observation in Fig. \ref{FA-KES}, that false claims, in this data set, have a higher intensity of “fear” compared to true ones. 

Overall, the emotional language analysis of false and true claims on different topics provides evidence supporting our hypotheses that the intention of the claims’ authors to incite emotion in their audience and the topic of the claims are inter-dependent and are influenced by the credibility of the claims.

\begin{table*}[t]
\begin{center}
\caption{Structural model results of hypothesis 2.}\label{Hyp2} 
\begin{tabular}{lllll}
\toprule
\multirow{2}{*}{Hypothesised Relationship}         & 
\multicolumn{4}{@{}c@{}}{Path Coefficient($t$-value)}
\\ 
\cmidrule{2-5}
&FA-KES & FakeHealth  & CoAID & Co-inform \\ 
\midrule

Claim Credibility $\,\to\,$ Claim Anger & 0.20 (6.82)***  & 0.07 (2.70)**  & -0.32 (8.37)***  & -0.21 (4.29)***\\

Claim Credibility $\,\to\,$ Claim Fear   & -0.79(23.65)*** & 0.05 (1.63)ns   & -0.30 (6.53)***  & 0.16 (3.95)*** \\ 

Claim Credibility $\,\to\,$ Claim Sadness  & 0.54 (16.10)*** & 0.27 (10.24)***   & 0.14 (3.82)***   & -0.08 (1.16)ns \\ 

Claim Credibility $\,\to\,$ Claim Happiness & 0.06 (1.77)ns  & -0.34 (15.03)***   & 0.28 (3.62)*** & 0.02 (0.43)ns \\ 

Claim Credibility $\,\to\,$ Claim Neutral   & 0.21 (4.12)***  & 0.35 (8.64)***   & 0.15 (3.57)*** & 0.19 (2.61)** \\
\bottomrule
\end{tabular}
\\\small{Note: * Significant at $p<0.05$, ** at $p<0.01$, and *** at $p<0.001$.}
\end{center}
\end{table*}


\subsubsection{Social-level Analysis}
This section evaluates Hypotheses 3 and 4 at the social-level, to understand the influence of emotions expressed in the claims and also credibility of the claims on social media users' response posts.

\begin{figure}[t]
  \includegraphics[width=\linewidth]{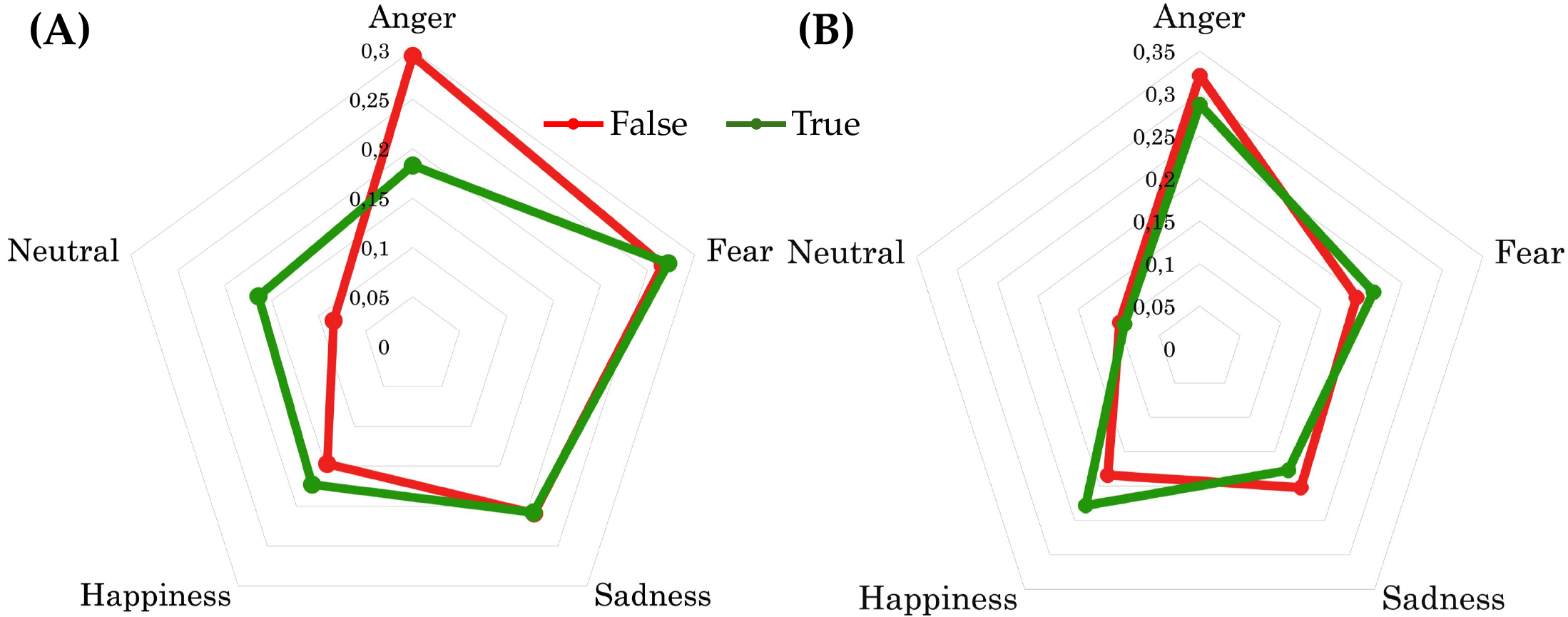}
  \caption{The average of emotion scores in replies to true and false claims in CoAID (A) and Co-inform (B) data sets.}\label{CoAID-Coinform-Replies}
\end{figure}

Fig. \ref{CoAID-Coinform-Replies} provides a comparison between the aroused emotions in users' responses to the claims on CoAID (chart A) and Co-inform (chart B) data sets by considering the claims' credibility. 
In corresponding replies to COVID-19 claims, the reflection of the emotions expressed in the claims (see Fig. \ref{CoAID-Coinform-Claims} chart A) can be seen in users' responses. Emotion analysis of users' response posts to false claims related to COVID-19 and the global pandemic shows users' reactions, on average, have a higher intensity of fear and anger than responses to true claims.
In the Co-inform data set, a similar distribution of emotion scores has been observed in users' responses and in the claims (see Fig. \ref{CoAID-Coinform-Claims} chart B), and users' dominant emotional reaction in the responses is anger which is also the major emotion in the claims.

\begin{figure}[t]
\begin{center}
  \includegraphics[width=\linewidth]{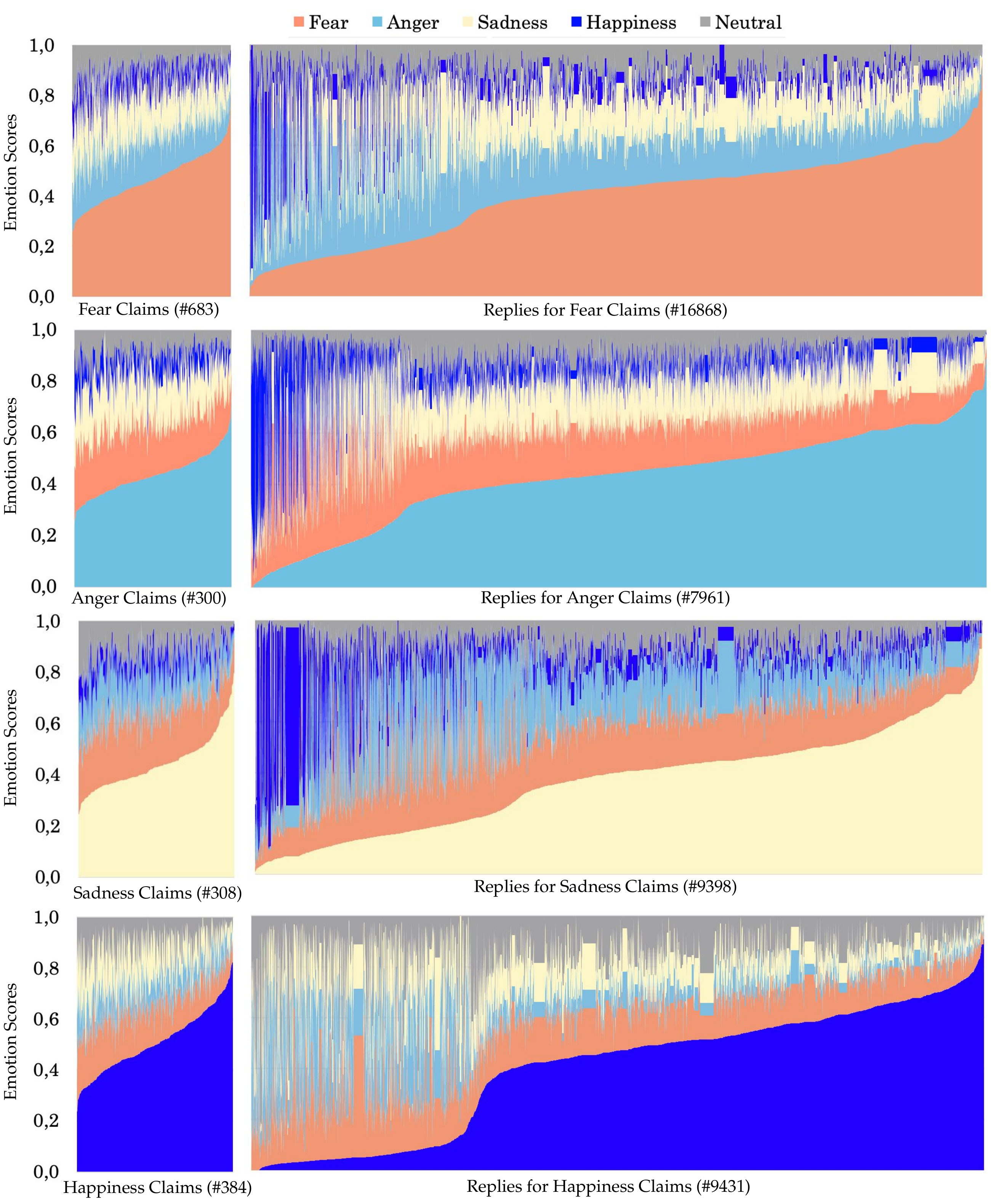}
  \caption{Emotional pattern diagram in the claims (left) and their related replies (right) in CoAID data set.}\label{CoAID2}
\end{center}
\end{figure}

To have a closer look at the effect of emotions in the claims related to COVID-19 and their audiences' responses on social media, we partitioned the claims based on the main emotions in them (i.e., fear, anger, sadness, and happiness). Fig. \ref{CoAID2} shows emotional pattern diagrams in claims with the highest fear, anger, sadness, and happiness scores and users' corresponding response posts on the right side in the CoAID data set. As is evident the emotional patterns in claims repeat themselves in users’ response posts, and the dominant emotion in the claims directly affected users’ emotions.

Tables \ref{Hyp3} and \ref{Hyp4} show the testing results of hypotheses 3 and 4 in CoAID and Co-inform data sets. As hypothesized, the results indicate the dominant emotion in a claim has a significant effect on users’ emotions expressed in their response posts, thus verifying H3. Likewise, as expected, the credibility of the claims significantly affects emotions of response posts, offering support for H4.

\begin{table}[t]
\begin{center}
\caption{Structural model results of hypothesis 3.}\label{Hyp3}
\tabcolsep=0.10cm
\resizebox{\linewidth}{!}{%
\begin{tabular}{lll}
\toprule
\multirow{2}{*}{Hypothesised Relationship}&  \multicolumn{2}{@{}c@{}}{Path Coefficient($t$-value)} \\ 
\cmidrule{2-3}
 &  CoAID &  Co-inform
\\
\midrule
Claim Anger $\,\to\,$ Replies Anger &  0.51 (11.19)***    &  0.44 (11.47)*** \\ 
Claim Fear $\,\to\,$ Replies Fear & 0.57 (15.72)***  & 0.43 (9.40)*** \\ 
Claim Sadness $\,\to\,$ Replies Sadness & 0.57 (12.71)***  & 0.39 (8.30)*** \\ 
Claim Happiness $\,\to\,$ Replies Happiness &  0.58 (12.98)***  & 0.38 (8.03)*** \\ 
Claim Neutral $\,\to\,$ Replies Neutral   & 0.32 (6.83)***  & 0.35 (5.38)***   \\
\bottomrule
\end{tabular}
}
\\\small{Note: * Significant at $p<0.05$, ** at $p<0.01$, and *** at $p<0.001$.}
\end{center}
\end{table}

\begin{table}[t]
\begin{center}
\caption{Structural model results of hypothesis 4.}\label{Hyp4}
\tabcolsep=0.10cm
\resizebox{\linewidth}{!}{%
\begin{tabular}{lll}
\toprule
\multirow{2}{*}{Hypothesised Relationship}&  \multicolumn{2}{@{}c@{}}{Path Coefficient($t$-value)} \\
\cmidrule{2-3}
&  CoAID &  Co-inform
\\
\midrule

Claim Credibility $\,\to\,$ Replies Anger          & -0.13 (3.99)***  & -0.14 (3.75)*** \\ 
 Claim Credibility $\,\to\,$ Replies Fear & -0.11 (3.65)*** & 0.03 (2.09)* \\ 
Claim Credibility $\,\to\,$ Replies Sadness        &  0.08 (2.43)*                & -0.10 (2.05)*   \\  
Claim Credibility $\,\to\,$ Replies Happiness &  0.10 (2.92)**  & 0.01 (3.10)** \\ 
Claim Credibility $\,\to\,$ Replies Neutral  & 0.21 (5.14)***   & 0.08 (3.90)*** \\ 
\bottomrule
\end{tabular}}
\\\small{Note: * Significant at $p<0.05$, ** at $p<0.01$, and *** at $p<0.001$.}
\end{center}
\end{table}

\subsubsection{Behavioral-level Analysis}
\begin{figure}[t]
      \centering
  \includegraphics[width=\linewidth]{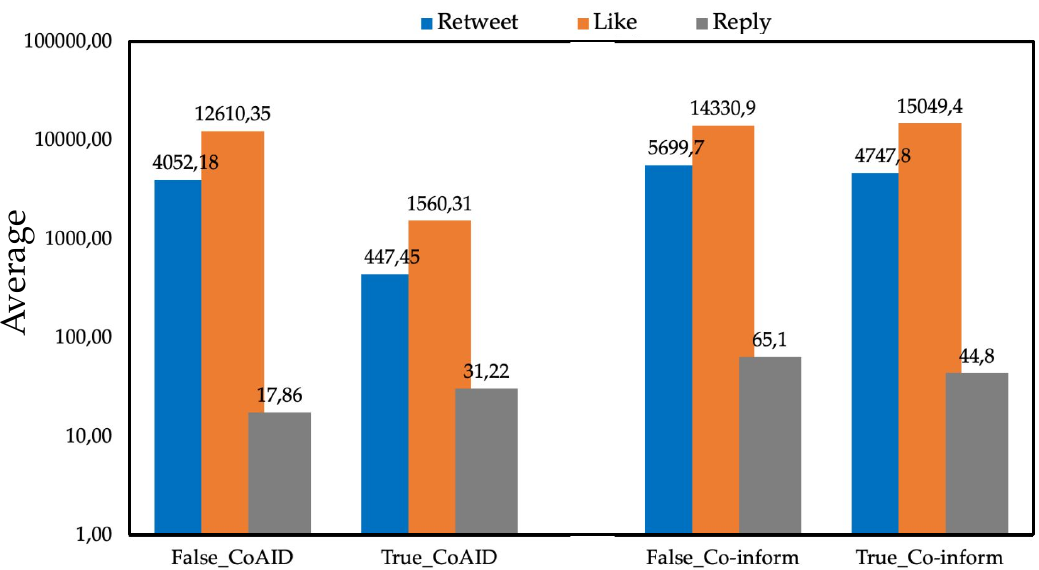}
  \caption{The average of retweets, likes and replies across true and false claims in the CoAID and Co-inform data sets.}\label{CoAID-Co-inform}
\end{figure}

This section evaluates Hypotheses 5, 6, and 7 at the behavioral-level, particularly, the impact of emotions expressed in false and true claims and corresponding users' response posts on their \textit{sharing} behavior in social media. Fig. \ref{CoAID-Co-inform} shows the average of retweets, likes, and replies in false and true claims in the CoAID and Co-inform data sets. In the CoAID data set, we can see that the average of retweets and likes in false claims is significantly higher than true ones, while, the average number of users' response posts to true claims is overtaking false claims. In the Co-inform data set, the tendency for behaviors like retweeting and responding is slightly higher for false claims than for true claims.

To determine the emotions that caused the highest levels of activity by users, we counted retweets, likes, and replies depending on dominating emotion. To this aim, the emotion with the highest score in a claim has been defined as its main emotion, and claims have been partitioned according to their main emotion. Tables \ref{CoAID-emotions} and \ref{CoInform-emotions} show the levels of activity in CoAID and Co-inform data sets depending on dominant emotions and credibility of claims. In the CoAID data set, the highest levels of activity were observed for claims expressing anger. Furthermore, negative emotions in false claims, especially anger and sadness, spurred higher numbers of retweets and likes than in true claims. In the claims expressing happiness, true claims have received higher retweets and likes from users than false claims. Similarly, in the Co-inform data set, in the claims carrying negative emotions, false claims caused higher retweets, like, and responses in users, and in the claims expressing happy emotions, true claims triggered a higher level of interaction between users. 

\begin{table}[t]
\begin{center}
\caption{Average of retweets, likes and replies separated by credibility and emotions in CoAID data set.}\label{CoAID-emotions}
\resizebox{\linewidth}{!}{%
\begin{tabular}{lcccc} 
\toprule
Emotion & Credibility (\#Claims) & Retweet & Like & Reply  \\ 
\midrule

\multirow{2}{*}{Anger} & False (265) & \textbf{11987.5} & \textbf{33383.0} & 19.2 \\ 
& True (35) & 2436.2 & 7943.9 & \textbf{82.2} \\ 
\cline{1-5}

\multirow{2}{*}{Fear} & False (398) & \textbf{1095.6} & \textbf{5389.6} & 15.8 \\ 
& True (285) & 728.0 & 2551.2 & \textbf{37.2} \\ 
\cline{1-5}

\multirow{2}{*}{Sadness} & False (89) & \textbf{2157.1} & \textbf{9229.6} & 4.1\\ 
& True (219) & 190.8 & 595.2 & \textbf{25.0} \\ 
\cline{1-5}

\multirow{2}{*}{Happiness} & False (156) & 177.7 & 734.2 & 9.0 \\ 
& True (228) & \textbf{365.7} & \textbf{1456.2} & \textbf{35.2} \\ 
\cline{1-5}

\multirow{2}{*}{Neutral} & False (38) & 24.1 & \textbf{49.3} & 5.4 \\ 
& True (179) & \textbf{30.1} & 47.8 & \textbf{15.7} \\

\bottomrule
\end{tabular}
}
\end{center}
\end{table}

\begin{table}[t]
\centering
\caption{Average of retweets, likes and replies separated by credibility and emotions in Co-inform data set.}\label{CoInform-emotions}
\resizebox{\linewidth}{!}{%
\begin{tabular}{lcccc} 
\toprule
Emotion & Credibility (\#Claims) & Retweet & Like & Reply  \\ 
\midrule

\multirow{2}{*}{Anger} & False (231) & \textbf{6236.6} & 14854.7 & \textbf{66.5} \\ 
 & True (152) & 5199.6 & \textbf{17424.5} & 54.3 \\ 
\cline{1-5}

\multirow{2}{*}{Fear} & False (39) & \textbf{9607.7} & \textbf{22772.4} & \textbf{65.4} \\ 
 & True (77) & 6355.1 & 12699.9 & 47.3 \\ 
\cline{1-5}

\multirow{2}{*}{Sadness} & False (67) & \textbf{5286.4} & \textbf{14519.1} & \textbf{61.2} \\ 
 & True (60) & 2956.4 & 8992.7 & 33.7 \\ 
\cline{1-5}

\multirow{2}{*}{Happiness} & False (84) & 2829.9 & 8785.9 & \textbf{61.0} \\ 
 & True (97) & \textbf{5951.6} & \textbf{23428.0} & 40.9 \\ 
\cline{1-5}

\multirow{2}{*}{Neutral} & False (24) & \textbf{5379.0} & \textbf{14454.0} & \textbf{69.5} \\ 
 & True (53) & 941.6 & 3173.5 & 30.1 \\
\bottomrule
\end{tabular}
}
\end{table}

The path coefficient, $t$-values, and significance, for hypotheses 5, 6, and 7 testing results in CoAID and Co-inform data sets are tabulated in Table \ref{Hyp5-6-7}. In the CoAID data set, the results indicate that the level of anger ($\beta = 0.44$, $p < 0.001$), fear ($\beta=0.22$, $p<0.05$), and sadness ($\beta = 0.28$, $p < 0.01$) in users' replies toward the claims have a significant effect on the sharing of claims. Likewise, in the Co-inform data set, negative emotions (anger, fear, and sadness) in users' replies have a significant effect on the sharing of claims. Thus, hypothesis 5  is supported. Meanwhile, the level of happiness in users' replies to claims, does not have a significant effect on the sharing behavior; thus, hypothesis H6 was not supported. In addition, neutrality in users’ replies does not affect sharing behavior ($p > 0.05$). Moreover, our analysis shows that the credibility of the claims has a negative direct effect on sharing of questionable claims ($\beta = -0.27$; $p < 0.001$) and ($\beta = -0.05$;  $p < 0.01$) in the CoAID and Co-inform data set, respectively. Therefore, hypothesis 7 is supported, meaning that claims' credibility significantly influences sharing them. The PLS results show that the variance explained for sharing behavior was approximately 52\% and 41\%, in CoAID and Co-inform data sets, respectively. Overall, the results provide evidence supporting our hypothesised model.

\begin{table}[t]
\centering
\caption{Structural model results of hypotheses 5, 6, and 7.}\label{Hyp5-6-7}
\tabcolsep=0.07cm
\resizebox{\linewidth}{!}{%
\begin{tabular}{clll}
\toprule
\multirow{2}{*}{Hypothesis} &  \multirow{2}{*}{Hypothesised Relationship} &  \multicolumn{2}{c}{Path Coefficient($t$-value)} \\ 
\cmidrule{3-4}
 &  &  CoAID &  Co-inform \\
\midrule
\multirow{5}{*}{H5-H6} & Replies Anger $\to$ Retweet     & 0.44 (5.80)***               & 0.31 (5.34)***  \\ 
 & Replies Fear $\to$ Retweet      & 0.22 (1.99)*    & 0.38 (7.72)*** \\ 
 & Replies Sadness $\to$ Retweet   & 0.28 (2.99)** & 0.12 (2.71)** \\ 
 & Replies Happiness $\to$ Retweet & 0.08 (0.66)ns  & 0.11 (2.12)*  \\ 
 & Replies Neutral $\to$ Retweet  &  0.04 (0.34)ns & 0.07 (0.13)ns \\ 
\cline{1-4}

H7  &  Claim Credibility $\to$ Retweet & -0.27 (10.91)*** & -0.05 (2.60)** \\ 
\bottomrule
\end{tabular}
}
\\\small{Note: * Significant at $p<0.05$, ** at $p<0.01$, and *** at $p<0.001$.}
\end{table}

\section{Conclusion}
Studies that investigate the behavior of social media users with regard to misinformation indicate that emotions in false claims play a crucial role in spurring people's willingness to share information on social media. However, these studies do not explain how emotions enhance the potency of false claims, causing them to spread more widely than true claims. In this paper, we have presented an in-depth analysis of the effects of emotional framing in false and true claims from different topics on the users’ emotions expressed in replies and distorting their sharing behavior. The emotional language analysis of the claims showed that the distribution of emotions can be influenced by both the topic and the credibility of the claims. The deeper analysis of the results showed the reflection of the emotions expressed in the claims in users' emotional responses on social media. Furthermore, our diffusion analysis suggests that not only the topic of the claim itself but also the emotion related to the topic is an effective subtle driving factor in users' retweeting decision. These findings suggest that social media users need to be aware of the potential influence of emotional appeals and the credibility of claims when sharing information online. 

Additionally, we found that the ability of emotions to influence sharing behaviors on social media depends on the specific emotion that was aroused. False claims motivate the audience to retweet the claims via negative emotional appeals such as anger, fear, and sadness. While, in claims that arouse happiness in the audience, true claims caused more interaction among users than false claims. In the future, we aim to explore the relationship between emotional framing and emotional contagion in online discussions. We will investigate the emotional contagion changes in online discussions over time, and the emotional contagion effect on spreading truth and falsity.

\bibliographystyle{ACM-Reference-Format}
\bibliography{sample-base}

\end{document}